\documentclass{ws-procs975x65}

\begin{document}

\title{\Large Rotating Braneworld Black Holes}

\author{ALIKRAM  N. ALIEV\footnote{aliev@gursey.gov.tr}}

\address{Feza G\"ursey Institute, P.K. 6  \c Cengelk\" oy, 34684 Istanbul, Turkey}

\begin{abstract}
We present a  Kerr-Newman type stationary and axisymmetric
solution that describes rotating black holes with a \textit
{tidal} charge in the Randall-Sundrum braneworld. The tidal charge
appears as an imprint of nonlocal gravitational effects from the
bulk space. We also discuss the physical properties of these black
holes and their possible astrophysical appearance.
\end{abstract}
\bodymatter
\section{Introduction}

The braneworld idea is a revolutionary idea  to relate the
properties of higher dimensional gravity to the observable world
by direct probing of TeV-size mini black boles at  high energy
colliders . According to this idea  our observable Universe is a
slice, a "3-brane" in higher dimensional space \cite{add1, rs2}.
This in particular gives: (i) An elegant geometric resolution of
the hierarchy problem between the electroweak scale and the
fundamental scale of  quantum gravity, (ii) the large size of the
extra dimensions supports the weakness of Newtonian gravity on the
brane and makes it possible to lower the scale of quantum gravity
down to the electroweak interaction scale, (iii) the braneworld
model (RS2 model) also supports the properties of four-dimensional
Einstein gravity in low energy limit. In light of all this, it is
natural to assume the formation of black hole in the braneworld
due to gravitational collapse of matter trapped on the brane.

Several strategies have been discussed in the literature to
describe the braneworld black holes. First of all, it has been
argued that if the radius of the horizon of a black hole on the
brane is much smaller than the size of the extra dimensions
$(\,r_{+} \ll L \,)$, the black hole, to a good enough
approximation, can be described by the usual classical solutions
of higher dimensional vacuum Einstein equations. In the opposite
limit when ($\,r_{+} \gg L \,)$, the black hole becomes
effectively four-dimensional with a finite extension along the
extra dimensions. The first simple solution pertinent to the
latter case is based on the idea of a usual Schwarzschild metric
on the brane that would look like a {\textit black string}
solution from the point of view of an observer in the bulk
\cite{chr}. However, the black string solution exhibits curvature
singularities at infinite extension along the extra dimension.

We shall discuss another strategy namely, we shall specify the
metric form induced on the 3-brane assuming a Kerr-Schild ansatz
for it. With this ansatz the system of the effective gravitational
field equations on the brane \cite{sms,ae1} becomes closed and the
solution to this system turns out to be a Kerr-Newman type
stationary axisymmetric black hole which possesses a \textit
{tidal} charge instead of a usual \textit{electric} charge.
\section{The metric form on the 3-brane}
To describe a rotating black hole in the Randall-Sundrum scenario
we shall make a particular assumption about metric on the brane,
taking it to be of  the Kerr-Schild form
\begin{equation}
ds^2= \left(ds^2 \right)_{flat} +H\, (l_{i} dx^{i}) ^2 \, ,
\label{kerrschild1}
\end{equation}
where  $ H $ is an arbitrary scalar function and $\,l_{i}\,$ is a
null, geodesic vector field in both the flat and full metrics.
Earlier \cite{dmpr}, this type of strategy was emoloyed for a
static black hole localized on the brane. With the metric form
(\ref{kerrschild1}) the effective gravitational field equations on
the brane
\begin{equation}
R_{ij}=-E_{ij}\,, \label{breq2}
\end{equation}
where $ E_{ij} $  the traceless "electric part" of the
five-dimensional Weyl tensor, and the associated  constraint
equation
\begin{equation}
R=0\,\,, \label{ham1}
\end{equation}
admit the solution which in the usual Boyer-Lindquist coordinates
takes the form \cite{ae2}
\begin{eqnarray}
ds^2 &=& -\left(1 - \frac{2Mr - \beta}{\Sigma}\right)dt^2 -
 \frac{2 \,a\left(2Mr -\beta\right)}{\Sigma} \,\sin^2\theta \, dt\,d\phi
\nonumber\\
&& +\frac{\Sigma}{\Delta}\,dr^2\ + \Sigma \,d\theta^2 +
\left(r^2+a^2+ \frac{2Mr-\beta}{\Sigma}\,\,a^2 \sin ^2\theta
\right)\sin^2\theta \, d\phi^2\, \,,\label{kn}
\end{eqnarray}
where
\begin{eqnarray}
\Delta &= & r^2 + a^2 - 2Mr + \beta\,\,,~~~~~\Sigma = r ^2 + a ^2
\cos ^2 \theta \,\,.\label{delta}
\end{eqnarray}
We see that that this metric looks exactly like the Kerr-Newman
solution in general relativity, in which  the square of the
electric charge  is "superceded" by a tidal charge parameter
$\,\beta\,$.  The Coulomb-type nature of the tidal charge is
verified by calculating  the components of the tensor $\,
E_{ij}\,$  through equation (\ref{breq2}). Therefore one can think
of it as carrying the imprints of nonlocal gravitational effects
from the bulk space. Furthermore, the tidal charge  may take on
both \textit{positive} and \textit{negative} values.
\section{Major Features}
In complete analogy to the Kerr-Newman solution in general
relativity, the metric (\ref{kn}) possesses two major features:
The event horizon structure and the existence of a  \textit{static
limit} surface, the ergosphere. The event horizon is a null
surface determined by the largest root of the equation
$\,\Delta=0\,$. We have
\begin{equation}
r_{+}= M + \sqrt{M^2 - a^2 - \beta}\, \label{horizon1}
\end{equation}
The horizon structure depends on the sign of the tidal charge. The
event horizon does exist provided that
\begin{equation}
 M^2 \geq a^2 +\beta\,.
\label{extreme}
\end{equation}
Thus, for the positive tidal charge we have the same horizon
structure as the usual Kerr-Newman solution. New interesting
features arise when the tidal charge is taken to be negative. For
$ \beta < 0 $ from equation (\ref{horizon1}) it follows that the
horizon radius
\begin{equation}
r_{+} \rightarrow \left(M + \sqrt{ -\beta}\right)\, > M \,
\label{greaterh}
\end{equation}
as $\,a\rightarrow M\,$. This is not allowed in the framework of
general relativity. From equations (\ref{horizon1}) and
(\ref{extreme}) it follows that for $ \beta < 0 $, the extreme
horizon $\,r_{+} = M \,$ corresponds to a black hole with rotation
parameter $\,a\,$ greater than its mass $\,M\,$\,. Thus, the bulk
effects on the brane \textit{may provide a mechanism for spinning
up the black hole  so that its rotation parameter exceeds its
mass}. Meanwhile, such a mechanism is impossible in general
relativity.

The static limit surface is determined by the equation
$\,g_{tt}=0\,$, the largest root of which gives the radius of the
\textit{ergosphere}
\begin{equation}
r_{0} = M + \sqrt{M^2 - a^2 \cos^2\theta - \beta}\,\,.
\label{ergo0}
\end{equation}
Clearly, this surface lies outside the event horizon coinciding
with it only at angles $\,\theta=0\,$ and $\,\theta=\pi\,$. The
negative tidal charge tends to extend the radius of the ergosphere
around the braneworld black hole, while the positive $\,\beta\,$
just as the usual electric charge in the Kerr-Newman solution,
plays the opposite role. For the extreme case, we find the radius
of the ergosphere within
\begin{equation}
M<r<M+\sin\theta\,\sqrt{M^2-\beta}\,\,. \label{ergob}
\end{equation}
We see that in astrophysical situations, the rotating braneworld
black holes with negative tidal charge are more energetic objects
in the sense of the extraction of the rotational energy from their
ergosphere.

\vfill
\end{document}